# Coordination of Interdependent Natural Gas and Electricity Systems Based on Information Gap Decision Theory

Farnaz Sohrabi [1], Farkhondeh Jabari [1], Behnam Mohammadi-Ivatloo [1*], Alireza Soroudi [2]

[1] Faculty of Electrical and Computer Engineering, University of Tabriz, 29 Bahman Blvd., Tabriz, Iran
[2] Energy Institute, University College Dublin, Dublin, Ireland
*bmohammadi@tabrizu.ac.ir

**Abstract:** The interactions of the natural gas network and the electricity system are increased by using gas-fired generation units, which use natural gas to produce electricity. There are various uncertainty sources such as the forced outage of generating units or market price fluctuations that affect the economic operation of both natural gas and electricity systems. This paper focuses on the steady-state formulation of the integrated natural gas transmission grid and electricity system by considering the uncertainty of electricity market price based on information gap decision theory. The higher and lower costs than the expected cost originated from the fluctuations of electricity market price are modelled by the robustness and opportunity functions, respectively. The objective is to minimize the cost of zone one while satisfying the constraints of two interdependent systems, which can obtain revenue from selling power to its connected zones in short-term scheduling. The capability of the proposed method is demonstrated by applying it on a 20-node natural gas network and IEEE RTS 24-bus. The proposed short-term coordination between natural gas and electricity infrastructures is solved and discussed.

## Nomenclature

| | |
|---|---|
| $h_{\acute{g}}$ | Coefficient for obtaining gas consumption of gas-fired generation units. |
| $B_c$ | Cost target for robustness function. |
| $B_w$ | Cost target for opportunity function. |
| $b_g$ | Cost coefficient of power generation unit $g$. |
| $c_k$ | Cost coefficient of gas supplier $k$. |
| $r_c$ | Critical value of the objective function. |
| $q$ | Decision variable in IGDT model. |
| $LE_{it}$ | Electrical demand of bus $i$ at time $t$. |
| $\tilde{\lambda}_t$ | Forecasted market price for interval $t$. |
| $\tilde{u}$ | Forecasted value of uncertain variable. |
| $LG_{mt}$ | Gas demand of node $m$ at time $t$. |
| $S_{kt}$ | Gas extraction from gas supplier $k$ at time $t$. |
| $\alpha$ | Horizon of the uncertain variable. |
| $g$ | Index for power generation units. |
| $\acute{g}$ | Index for gas-fired generation units. |
| $i,j$ | Index for network buses in zone one. |
| $k$ | Index for natural gas suppliers. |
| $m,n$ | Index for nodes in natural gas transmission. |
| $t$ | Index for hours. |
| $w$ | Index for wind power generation units. |
| $z$ | Index for network buses in zones two and three. |
| $\hat{\alpha}(q,r_c)$ | Information-gap robustness function. |
| $\hat{\beta}(q,r_w)$ | Information-gap opportunity function. |
| $\lambda_t$ | Market clearing price at time $t$. |
| $P_g^{max}$ | Maximum power generation limit of unit $g$. |
| $P_{ij}^{max}$ | Maximum power flow limits between bus $i$ & $j$. |
| $P_{iz}^{max}$ | Maximum power flow limits between bus $i$ & $z$. |
| $P_w^{max}$ | Maximum power generation limit of unit $w$. |
| $S_k^{max}$ | Maximum limit of gas extraction from well $k$. |
| $\pi_m^{max}$ | Maximum limit of pressure at node $m$. |
| $P_g^{min}$ | Minimum power generation limit of unit $g$. |
| $P_w^{min}$ | Minimum power generation limit of unit $w$. |
| $S_k^{min}$ | Minimum limit of gas extraction from well $k$. |
| $\pi_m^{min}$ | Minimum limit of pressure at node $m$. |
| $f_{mnt}$ | Natural gas flow from node $m$ to $n$ at time $t$. |
| $\pi_{mt}$ | Node pressure. |
| $P_{ijt}$ | Power flow between bus $i$ and $j$ at time $t$. |
| $P_{izt}$ | Power flow between bus $i$ and $z$ at time $t$. |
| $P_{gt}$ | Power generation of unit $g$ at time $t$. |
| $P_{\acute{g}t}$ | Power generation of gas-fired unit $\acute{g}$ at time $t$. |
| $P_{wt}$ | Power generation of unit $w$ at time $t$. |
| $R_g^{dn}$ | Ramp-down limit of unit $g$. |
| $R_g^{up}$ | Ramp-up limit of unit $g$. |
| $\Omega_A$ | Set of gas pipelines in zone one. |
| $\Omega_B$ | Set of network buses in zone one. |
| $\Omega_G$ | Set of all power generation units. |
| $\Omega_G^i$ | Set of all power generation units connected to bus $i$. |
| $\Omega_l$ | Set of network branches. |
| $\Omega_l^i$ | Set of all buses connected to bus $i$. |
| $\Omega_N$ | Set of gas nodes. |
| $\Omega_K^m$ | Set of all natural gas suppliers connected to node $m$. |
| $\Omega_G^m$ | Set of all gas-fired units connected to node $m$. |
| $r_w$ | Target value of the objective function. |
| $Cost$ | Total cost. |
| $x_{ij}$ | Transmission line reactance between buses $i$ and $j$. |
| $u$ | Uncertain parameter in IGDT model. |
| $U(\alpha,\tilde{u})$ | Uncertainty model in IGDT method. |
| $\delta_{it}$ | Voltage angle in bus $i$. |
| $C_{mn}$ | Weymouth constant. |

## 1. Introduction

Natural gas (NG) and electricity are two important energy sources. The use of NG to generate electricity via natural gas-fired generating units has increased the interdependency between NG and electricity energy sources [1]. Due to the interactions between NG and electricity systems, their economy and reliability are affected by each other. Thus, the operating cost of gas-fired units changes if the NG's price fluctuates [2-4]. By 2030, the usage of NG in power generation is expected to increase by 230% [5]. In North America, gas-fired units were anticipated to supply more than half of the peak electricity demand by 2015 [4]. The integrated energy systems (IES) show more advantages compared to the independent ones [6]. Combined cycle power plants and gas-fired units have lower capital cost investment, operation flexibility, lower emission and better economic efficiency compared to the conventional coal plants [7].

Considering the above-mentioned benefits for energy system integration, the coordination between NG and electricity has obtained significant attention in many research works. The interdependency between NG and electricity systems discussed in [8-13]. In [8], the Monte Carlo simulation was applied to represent the coordinated stochastic model of hourly economic demand response of electric power systems as well as the constraints of NG transmission by considering random errors in forecasting the day-ahead hourly loads, random outages of transmission lines and generating units. The optimal short-term operation of both hydrothermal systems and NG network was studied in [9] by assuming the constraints at the hydrothermal system, NG pipeline, extraction, and storage operation. In [10], the Lagrangian relaxation (LR) was implemented to relax the coupling constraints of the coordinated scheduling of interdependent NG transmission systems and electric power by the goal of minimizing the social cost. Reference [11] used Newton's method to analyse the integrated electricity and NG systems in steady-state mode by assuming the temperature effect in the NG network operation in addition to a distributed slack node technique in the power system. Due to the linkage between NG and power systems via gas-fired power plants, the NG transmission can affect power transmission in terms of security and economics. Thus Ref. [12] investigated the short-term security-constrained unit commitment (SCUC) with NG transmission constraints to minimize the operating cost and determine the hourly unit commitment (UC) and dispatch while meeting the NG and electricity constraints. Munoz et al. [13] proposed a two-phase nonlinear optimization method to model NG for power system reliability studies.

The coordination of NG and electricity systems can be investigated in two groups, uncertainty–neutral or uncertainty-constrained. In the first group, it is considered that the accurate value of the uncertain parameter is available. However, in the second group, the variation of the uncertain parameter is unknown. The uncertainty in the coordination of the NG and electricity systems has been studied in [14-17]. The authors in [14] investigated the effect of high wind penetration on the Great Britain gas network by considering the distinct feature of gas and electric power flows, ramping characteristics of variant power plants and gas support facilities like storage and compressors. The midterm stochastic SCUC is applied in [15] to coordinate NG and hydro systems for incorporating high wind integration. A deterministic and multi-stage stochastic programming models is proposed in [16] to study integrated NG and electricity systems in Great Britain with wind uncertainty. An interval optimization-based operating strategy for integrated gas and electricity energy systems is proposed in [17] to improve the system operation by taking into account the demand response and wind power uncertainty.

There are different methods to model uncertainties. Some models like fuzzy or stochastic model need membership function or probability density functions are needed to quantify the uncertainties [18, 19]. IGDT was developed as an alternative model to decide under severe uncertainty. The sever uncertainty refers to a situation where no membership function or probability density function is available about the uncertain input parameter. To explain more, IGDT models the gap between what is known and what is unknown for the decision maker to make informed decisions by recognizing opportunities and risks. Uncertainties may be damaging or beneficial, which leads to lower or higher profits, respectively. These two conflicting issues are investigated in IGDT model using two immunity functions of robustness and opportunity.

One way to model uncertainty is information gap decision theory (IGDT) which was proposed by Ben Haim [20]. To model the uncertainties of market prices, Ref. [21] proposed a non-probabilistic information-gap model for risk-neutral, opportunity seeker and risk-averse generation companies (GenCos) for maximizing the gained profit in short-term scheduling. Based on IGDT model, a decision making framework was discussed in Ref. [22] to help the distribution network operators (DNOs) in choosing the supplying resources comprised of Distributed Generations (DGs), the pool market and the bilateral contracts to satisfy customers' demand. In this model, the uncertain parameters such as demand of each bus, the electricity price in the pool market and the decisions of DG investors were taken into account. An IGDT-based risk-constrained bidding and offering strategy in the day-ahead energy markets for a merchant compressed air energy storage (CAES) plant was modelled in [23] to manage the risk of price forecast errors due to price uncertainty.

A robust SCUC model was presented in Ref. [24] to increase the operational reliability of integrated NG and electricity systems against the possible transmission line outages. Based on the Ref. [25], robust optimization (RO) and IGDT are similar, and both are classified in the interval optimization methods. In both methods, the model of the uncertainty formulation and the risk hedging of uncertain parameter look like each other. However; the big difference between RO and IGDT is their inputs, which differentiates them significantly in applications and makes RO less comprehensible and user-friendly than IGDT from a financial viewpoint. In RO, the confidence interval boundaries are the optimization input; however, in IGDT, the desired amount of cost function is the input. In RO, the boundaries of the uncertain parameter are the input parameters for calculating the guaranteed profit; but, in IGDT, the user sets the guaranteed profit to maximise the confidence interval. Moreover, the opportunistic optimization can be modeled in IGDT, which is impossible in RO.

Different origins of uncertainty in gas-power nexus can be identified. In [26], the price uncertainty was studied and according to the market structure and forecasting method, the price forecasting errors can be tolerated from 5% to 36%. Hence, in this paper, according to the major effect of the electricity market price on overall cost, the coordination between NG infrastructure and the electricity network is proposed by modelling the price uncertainty based on IGDT method. The objective function of the integrated NG and electrical system is to minimize the cost in zone one for the opportunity-seeker and risk-averse power generation units over a multi-hour operation period. It should be noted that zone one is considered as the price-taker zone that means that the operation of this zone does not significantly influence the clearing price of the imported electricity from neighbour zones. The contributions of this paper are described as follows:

a) The performance of integrated NG and electricity systems in zone one is optimized by considering the electricity price uncertainty for selling power to its connected zones (i.e., zones two and three).

b) The IGDT approach is employed to model the electricity price uncertainty.

c) Robustness function of IGDT approach is used to obtain the robust strategy in operation of integrated NG and electricity systems.

d) Opportunity function of IGDT approach is implemented to gain opportunity strategy

The paper is structured into these sections: Section II provides the problem formulation of the integrated NG and electricity systems. The mathematical formulation of IGDT method is investigated in this section. Section III explains the solution methodology. The simulation results are discussed in section IV. In section V, the conclusion is made.

## 2. Mathematical formulation

### 2.1. Model of integrated natural gas and electricity systems

This paper concentrates on the steady-state analyses of the electricity and the NG systems. NG network is a complicated non-linear system, which is comprised of pipelines, compressors and wells. NG system transports gas from wells to end users via pipelines. The steady state model of the NG network is given in [12]. In this study, the DC optimal power flow is used to model the electricity network. The coordination of interdependent NG and electricity systems is modelled in [30]. The topology of the test system is shown in Fig. 1 [30].

The objection function is to minimize the cost of zone one which can gain revenue from selling power to its connected zones over the optimization period. The first and second terms of (1) show the total expenses of electricity and NG systems in zone one, respectively. The third term is the obtained revenue of zone one from selling power to zones two and three through power transmission lines.

$$Cost = \sum_{gt} P_{gt} b_g + \sum_{kt} S_{kt} c_k - \sum_t \left[ \sum_{iz \in \Omega_l} P_{izt} \lambda_t \right] \quad (1)$$

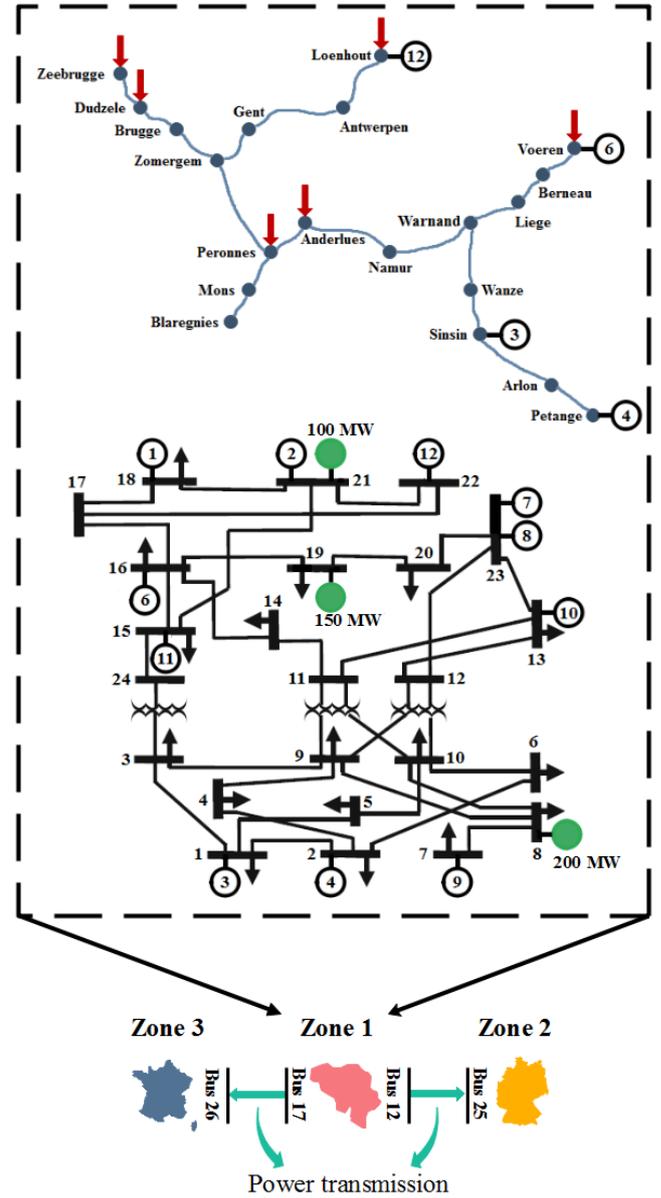

**Fig. 1.** *Topology of the system.*

The total injected energy should be equal to the total withdrawn energy in each bus in an electricity network and each node in a NG system. The power balance and gas balance are shown in (2) and (3), respectively. The gas-fired power plants link the electricity power system and the NG network. Gas-fired power plant produces power for the electricity system so it is a supplier for the power system. However, it consumes NG thus it is a load for the NG system.

$$\sum_{w \in \Omega_G^i} P_{wt} + \sum_{g \in \Omega_G^i} P_{gt} - LE_{it} = \sum_{j \in \Omega_l^i} P_{ijt} + \sum_{z \in \Omega_l^i} P_{izt} ; \forall i \in \Omega_B, \forall t \quad (2)$$

$$\sum_{n|(m,n) \in \Omega_A} f_{mnt} = \sum_{n|(n,m) \in \Omega_A} f_{nmt} + S_{kt} - LG_{mt} - \sum_{\acute{g} \in \Omega_G^m} P_{\acute{g}t} h_{\acute{g}} ; \forall m \in \Omega_N, \forall k \in \Omega_K^m, \forall t \quad (3)$$

The DC optimal power flow between connected buses in the electricity network is calculated by line impedance and bus angles in zone one and connected zones as (4) and (5), respectively.

$$P_{ijt} = \frac{\delta_{it} - \delta_{jt}}{x_{ij}}; \forall ij \in \Omega_l, \forall t \quad (4)$$

$$P_{izt} = \frac{\delta_{it} - \delta_{zt}}{x_{iz}}; \forall iz \in \Omega_l, \forall t \quad (5)$$

The power production of each unit in power system is limited by upper and lower generating bounds of unit as follows:
$P_g^{min} \leq P_{gt} \leq P_g^{max}; \forall g \in \Omega_G, \forall t$ (6)
$P_w^{min} \leq P_{wt} \leq P_w^{max}; \forall w \in \Omega_G, \forall t$ (7)

The rate of generated output power should be modified in an admissible range. The ramp-up and ramp-down limits of power generation units are mathematically written in (8) and (9).
$P_{g(t+1)} - P_{g,t} \leq R_g^{up}; \forall g \in \Omega_G, \forall t$ (8)
$P_{g(t-1)} - P_{g,t} \leq R_g^{dn}; \forall g \in \Omega_G, \forall t$ (9)

The maximum and minimum limitations of power flow in zone one and connected zones are as (10) and (11).
$-P_{ij}^{max} \leq P_{ijt} \leq P_{ij}^{max}; \forall ij \in \Omega_l, \forall t$ (10)
$0 \leq P_{izt} \leq P_{iz}^{max}; \forall iz \in \Omega_l, \forall t$ (11)

The nodal pressures and flow rates are the variables of NG network which are determined by solving the steady state NG problem. The NG flow through a pipeline between gas node $m$ and $n$ without compressor is a non-linear function of the technical parameters and the nodal pressures as follows:
$f_{mnt} = \text{sgn}(\pi_{mt}, \pi_{nt}). C_{mn}\sqrt{|\pi_{mt}^2 - \pi_{nt}^2|}$ (12)
$\text{sgn}(\pi_{mt}, \pi_{nt}) = \begin{cases} +1 & \pi_{mt} \geq \pi_{nt} \\ -1 & \pi_{mt} < \pi_{nt} \end{cases}$ (13)

where $C_{mn}$ is the Weymouth constant and depends on pipeline characteristics [27]. The higher pressure determines the direction of gas flow meaning that the gas flows from a node with high pressure to low pressure gas node. The NG flow through a pipeline between gas node $m$ and $n$ with compressor is stated as (14).
$f_{mnt} \geq \text{sgn}(\pi_{mt}, \pi_{nt}). C_{mn}\sqrt{|\pi_{mt}^2 - \pi_{nt}^2|}$ (14)

Through the pipeline, the pressure decreases, thus the compressors are used to increase pressure level and transmission efficiency. In other words, they act like transformers in a power system.

The pressure at each node is restricted by its upper and lower bounds as (15).
$\pi_m^{min} \leq \pi_{mt} \leq \pi_m^{max}; \forall m, \forall t$ (15)

The maximum and minimum limits for gas extraction from wells are modelled as (16).
$S_k^{min} \leq S_{kt} \leq S_k^{max}; \forall k, \forall t$ (16)

### 2.2. Information gap decision theory

In the IGDT method, the horizon of uncertainty is maximized while the specific requirement is guaranteed as investigated in [20, 21]. The IGDT method has two different performances comprising robustness function and opportunity function, which are risk-averse and opportunity-seeker terms, respectively. In the IGDT method, the cost target is selected as the input parameter of the optimization problem and set by the user. Then, the maximum length of the confidence interval related to the uncertain parameter would be calculated. In both robustness and opportunity models, the electricity market price is selected as the uncertain parameter. Then, the maximum value of the market price reduction is found in a way that the cost is equal to or smaller than the given cost target for robustness function. In the opportunity seeker decision-making process, the minimum value of the market price increase is optimally determined in a way that the cost is equal to or smaller than the given cost target for the opportunity function. The robustness and opportunity functions are shown in (18) and (19), respectively.

*2.2.1. Uncertainty model:* Uncertainty model in IGDT is applied to consider the gap between real amount and the predicted value of the uncertain parameter.
$U(\alpha, \tilde{u}) = \{u: |u_t - \tilde{u}_t| \leq \alpha \phi_t\}, \alpha \geq 0$ (17)

where, $\phi(t)$ is the envelope shape of uncertainty and can be in different forms. In this paper, $\phi_t$ is considered as the nominal value of uncertain variable.

*2.2.2. Robustness function:* The robustness function in IGDT states the maximum uncertainty of the decision variable $q$ while the critical value of the objective function $r_c$ is satisfied.
$\hat{\alpha}(q, r_c) = \max_{\alpha}\{\alpha : \text{Objective function} \leq r_c\}$ (18)

*2.2.3. Opportunity function:* The opportunity function in IGDT describes the minimum uncertainty of the decision variable $q$ while the target value of the objective function $r_w$ is satisfied.
$\hat{\beta}(q, r_w) = \min_{\alpha}\{\alpha : \text{Objective function} \leq r_w\}$ (19)

## 3. Integrated natural gas and electricity based IGDT method

The IGDT-based formulation for integrated NG and electricity is proposed in this section. The market price of the electricity power $\lambda$ is the uncertain parameter. The forecasted values of $\lambda$ which are denoted by $\tilde{\lambda}$ are supposed to be available. The objective function is to minimize the cost of zone one for the total operation period. Based on the background in previous section, the uncertainty model of market price is written as (20).
$U(\alpha, \tilde{\lambda}) = \left\{\lambda : \frac{|\lambda_t - \tilde{\lambda}_t|}{\tilde{\lambda}_t} \leq \alpha\right\}; \alpha \geq 0$ (20)

Relation (20) is simplified to inequality constraints (21) and (22).
$\lambda_t \leq (1 + \alpha)\tilde{\lambda}_t$ (21)
$\lambda_t \geq (1 - \alpha)\tilde{\lambda}_t$ (22)

The robustness function is scheduled in a way that the maximum cost of zone one is not higher than a specified cost target while the unfavorable deviations of market price occur. The cost target for robustness function $r_c$ which is shown by $B_c$ is gained as:
$B_c = (1 + \sigma)B_0$ (23)

where $B_0$ is the base cost without applying IGDT. It means that $B_0$ is the expected minimum cost based on the forecasted market prices where the risk is not taken into account. The cost deviation factor $\sigma$ is used to increase the base cost. It is obvious that the maximum cost is happened when the market clearing price is lower than the forecasted price so $\lambda$ is as follows:
$\lambda_t = (1 - \alpha)\tilde{\lambda}_t$ (24)

To explain more, the risk-averse decision maker desires to schedule in a way to be immune against the high cost of zone one due to the unfavorable deviation of the market prices in selling power to the connected zone, from the forecasted values. Hence, the robustness function can be modelled as:
$\hat{\alpha}(P, B_c) = \max \alpha$ (25)
Subject to:
$\sum_{gt} P_{gt} b_g + \sum_{kt} S_{kt} c_k - \sum_t [\sum_{iz \in \Omega_l} P_{izt}(1 - \alpha)\tilde{\lambda}_t] \leq B_c$

The robustness function, which has a risk-averse characteristic, states that the cost is not higher than $B_c$ when the maximum deviation of the forecasted price occurs and the price of market declines to $(1-\alpha)\tilde{\lambda}(t)$.

The market prices higher than the forecasted ones result in the cost decrease. The opportunity function, which has a risk-seeker characteristic is scheduled to benefit from this desirable deviation of market price. The cost target for opportunity function $r_w$ which is shown by $B_w$ is gained as:

$$B_w = (1-\rho)B_0 \qquad (26)$$

where $\rho$ is a desired deviation parameter to show a lower cost from $B_0$. The minimum cost occurs when the price of clearing market becomes higher than the forecasted price thus $\lambda$ is as follows:

$$\lambda_t = (1+\alpha)\tilde{\lambda}_t \qquad (27)$$

The opportunity function which models the possible low cost of zone one for the risk-seeker decision maker can be expressed as:

$$\hat{\beta}(P, B_w) = \min \alpha \qquad (28)$$

Subject to:
$$\sum_{gt} P_{gt} b_g + \sum_{kt} S_{kt} c_k - \sum_t \left[ \sum_{iz \in \Omega_l} P_{izt}(1+\alpha)\tilde{\lambda}_t \right] \leq B_w$$

The opportunity function states that the cost is lower than $B_w$ if the minimum desired deviation $\alpha$ occurs and increases the price to $(1+\alpha)\tilde{\lambda}_t$. In other words, if the future electricity prices favorably deviate from $\tilde{\lambda}_t$ by $\hat{\beta}$, a lower cost of $B_w$ for zone one may be achieved. It should be mentioned that $\hat{\beta}$ is the minimum required electricity price deviation that makes $B_w$ achievable.

## 4. Simulation results and discussion

The numerical simulations of coordination between interdependent NG and electricity systems based on IGDT help us to study the proposed procedure. The discussed problem in (1)-(28) is solved in Generalized Algebraic Modeling Systems (GAMS) software [28, 29] using the code provided in [30]. In the robust mode of IGDT-based NLP problem, the decision maker desires that the cost of the integrated system calculated as the fuel cost of thermal units plus the gas extraction cost minus the revenue obtained from selling energy is minimized as low as the target cost, $B_c$. Meanwhile, the maximum percentage of the electricity price decrease can be found in the main objective function. Moreover, the operational constraints of the interconnected electricity and NG systems should be satisfied. It is supposed that the cost target for robustness function is calculated as (23) and is equal to $B_c = (1+\sigma)B_0$ in which, $B_0$ refers to the operating cost of the system before the implementation of IGDT strategy and equals to $423483. In base problem, without application of IGDT method, the operation cost of the gas and electricity grids is higher than the revenue obtained from the selling electricity to zones two and three. Hence, the cost is considered for calculations as obvious from (1). In the opportunistic model, the operator of the zone one desires that its cost is reduced as low as $B_w = (1-\rho)B_0$. Moreover, the minimum value of the electricity price increase, which results in the reduction of cost to $B_w$, is found as the main objective function. In other words, the decision maker determines that the cost of the interconnected systems is lower than or equal to the cost target $B_c$ in robust mode and the cost target $B_w$ in opportunity model. In the IGDT method, the cost which should be guaranteed is selected as the input parameter of the optimization problem and set by the user. Then, the maximum length of the confidence interval related to the electricity market price would be calculated. In both robustness and opportunistic modes, the electricity market price is selected as the uncertain parameter. Then, the maximum value of the market price decrease will be found in a way that the cost is equal to or less than the given cost target. In the risk-seeker decision making process, the minimum value of the market price increase will be optimally determined in a way that the cost is equal to or lower than the given cost target.

### 4.1. Data

The proposed methodology is applied to IEEE RTS 24-bus with 20-node NG system which is defined as zone one in this paper and the parameters of the test system can be found in [30]. The operation time of the simulation is considered as 24 hours. The IEEE RTS 24-bus includes 12 generation units and 34 branches. 4 out of 12 generation units are gas-fired units. Generator 3 at bus 1, generator 4 at bus 2, generator 6 at bus 16 and generator 12 at bus 22 are gas-fired units. The 20-node NG system includes 6 gas suppliers, 24 pipelines, 3 compressors and 9 gas loads. The Branch data for connected zones to zone one is given in Table 1.

**Table 1** Branch data for connected zones to zone one

| From | To | $x(pu)$ | Limit($MW$) |
|---|---|---|---|
| 12 | 25 | 0.0245 | 900 |
| 17 | 26 | 0.0108 | 1000 |

### 4.2. Results of the robust model

By solving the robust optimization problem, the optimum robustness function value $\hat{\alpha}$ for different amounts of deviation factor $\sigma$ from 0 to 0.9 is demonstrated in Fig. 2. Observe that $\sigma = 0$ corresponds to the risk-neutral case where the cost target for the robustness function $B_c$ is equal to the expected cost $B_0$ which is earned by solving the deterministic scheduling problem based on the forecasted prices $\tilde{\lambda}(t)$. As seen in this figure, by rising $\sigma$ the value of $\hat{\alpha}$ increases which means that zone one should pay higher cost in order to have a robust strategy. The cost target for the robustness function of the IGDT-based optimization problem is tabulated in Table 2. To explain more if the forecasted price $\tilde{\lambda}(t)$ decreases to $(1-\hat{\alpha})\tilde{\lambda}(t)$, it is guaranteed that the cost will not be higher than $B_c$.

**Table 2** Cost target for the robustness function

| $\sigma$ | $\hat{\alpha}$ | $B_c$ ($) |
|---|---|---|
| 0.0 | 0.000 | 423483 |
| 0.1 | 0.041 | 465832 |
| 0.2 | 0.081 | 508180 |
| 0.3 | 0.122 | 550528 |
| 0.4 | 0.164 | 592877 |
| 0.5 | 0.206 | 635225 |
| 0.6 | 0.251 | 677573 |
| 0.7 | 0.298 | 719922 |
| 0.8 | 0.357 | 762270 |

| 0.9 | 0.450 | 804619 |

For example, for σ=0.5 it is guaranteed that the cost will not be higher than $B_c = (1 + 0.5)B_0 = (1 + 0.5)\,423483 = \$\,635225$ if the maximum electricity price reduction at time *t* equals to 0.206 which means that electricity price can be reduced to (1-0.206)=0.794 of the forecasted price data, $\tilde{\lambda}(t)$.

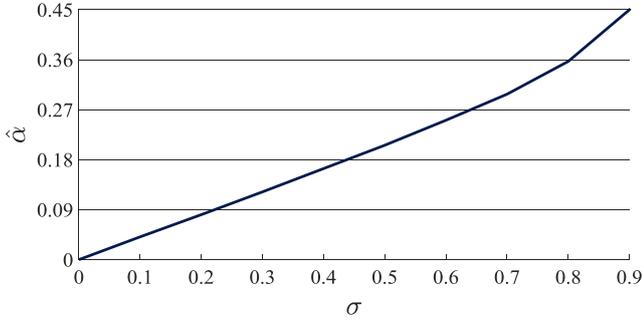

**Fig. 2.** *Optimum robustness function value versus cost deviation factor.*

The total transmitted power from zone one to its connected zones, the power generation and the gas consumption of zone one for different values of σ from 0 to 0.9 are written in Table 3. It is clear that by increasing σ the transmitted power are decreased which results in the lower power generation and gas consumption of zone one.

**Table 3** Output results for the robustness schedule

| σ | $P_{iz}(MW)$ | $P_g(MW)$ | $S_k(10^6 Sm^3)$ |
|---|---|---|---|
| 0.0 | 33453 | 78451 | 40.198 |
| 0.1 | 33228 | 78226 | 40.177 |
| 0.2 | 33088 | 78085 | 40.160 |
| 0.3 | 32654 | 77651 | 40.103 |
| 0.4 | 32041 | 77039 | 39.912 |
| 0.5 | 30693 | 75691 | 39.773 |
| 0.6 | 29333 | 74331 | 39.681 |
| 0.7 | 27253 | 72251 | 39.296 |
| 0.8 | 16665 | 61663 | 38.135 |
| 0.9 | 12397 | 57394 | 37.867 |

The hourly robust schedule of the transmitted power, two selected generation units 3 and 11 as well as the natural gas suppliers of Peronnes and Voeren under different cost deviation factors σ=0.08 and σ=0.6 are reported in Figs. 3-5 to investigate the effect of the different cost targets on the scheduling, respectively. The results are gained for $B_c = (1 + 0.08)B_0 = \$\,457362$ and $B_c = (1 + 0.6)B_0 = \$\,677573$. It is evident from Fig. 3 that the transmitted power is declined for the higher σ. As seen in Fig.4 for the higher σ, the output power of the generation units is decreased or at least does not change because an increase in σ results in higher $\hat{\alpha}$ and thus lower power selling price. It is obvious that a decrease in the electricity price results in lower power production of gas-fired units and thus the gas extraction from the gas wells is reduced as shown in Fig. 5.

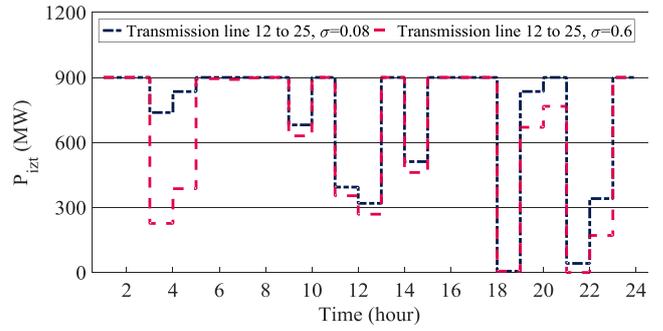
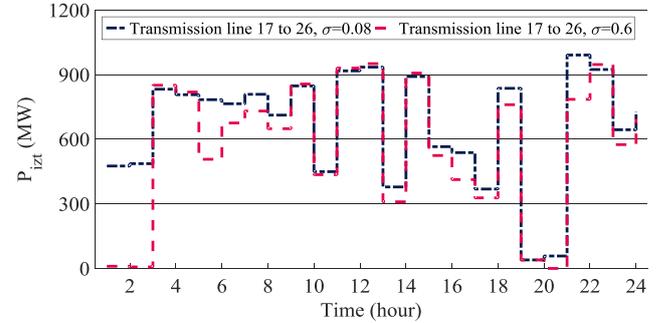

**Fig. 3.** *Robustness schedule of transmitted power for two different cost targets.*

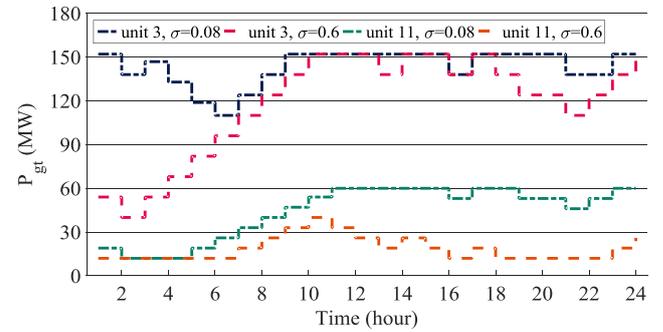

**Fig. 4.** *Robustness schedule of units 3 and 11 for two different cost targets.*

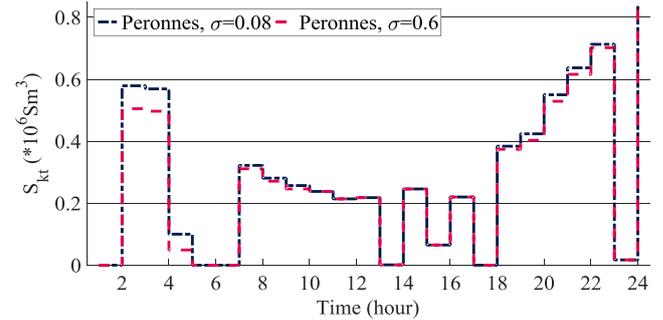
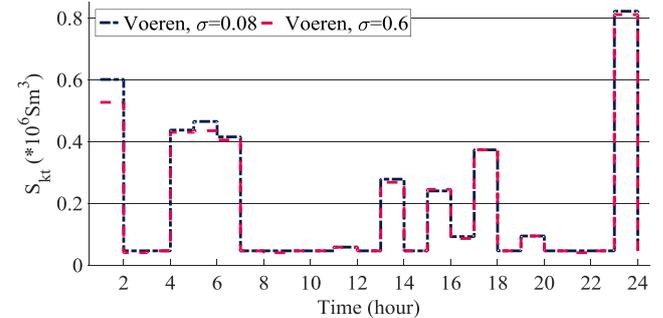

**Fig. 5.** *Robustness schedule of gas suppliers for two different cost targets.*

According to Figs. 3-5, by rising the cost deviation factor $\sigma$, the optimum schedule of the power generation units and gas suppliers changes in a way that the transmitted electrical power through the boundaries 12 to 25 and 17 to 26 as well as the obtained revenue from selling energy is decreased. Therefore, the cost of zone one is increased.

### 4.3. Results of the opportunity model

For analyzing the impact of modeling opportunity on the cost, the optimum opportunity function value $\hat{\beta}$ for different cost deviation factor from $\rho = 0$ to $\rho = 0.6$ is found and depicted in Fig. 6. It is obvious that an increase in $\rho$ leads to the decrement in the cost target for opportunity function $B_w = (1 - \rho)B_0$, which means that higher positive price spikes are required to gain lower cost. In simple words, if the forecasted price $\tilde{\lambda}(t)$ increases up to $(1 + \hat{\beta})\tilde{\lambda}(t)$, it is guaranteed that the cost of zone one will not be higher than $B_w$ where the variations of $\hat{\beta}$ versus $B_w$ are reported in Table 4.

**Table 4** Cost target for the opportunity function

| $\rho$ | $\hat{\beta}$ | $B_w$ ($) |
| --- | --- | --- |
| 0.0 | 0.000 | 423483 |
| 0.1 | 0.040 | 381135 |
| 0.2 | 0.081 | 338787 |
| 0.3 | 0.121 | 296438 |
| 0.4 | 0.161 | 254090 |
| 0.5 | 0.201 | 211742 |
| 0.6 | 0.241 | 169393 |

For instance, if the decision maker of zone one desires that the cost target for opportunity function be less than $ 296438, the electricity price at time *t* should be minimally increased to (1+0.121)=121.1% of the forecasted price $\tilde{\lambda}(t)$.

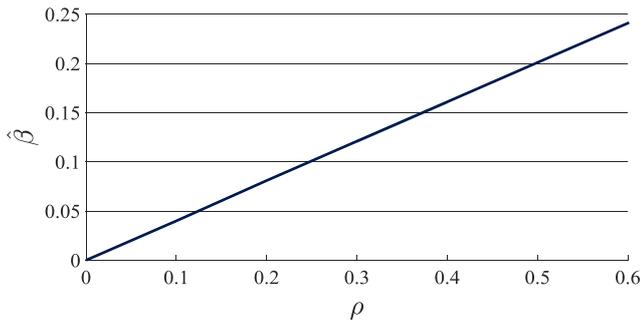

***Fig. 6.*** *Optimum Opportunity function value versus cost deviation factor.*

The total transmitted power from zone one to its connected zones, power generation and gas consumption of zone one for different values of $\rho$ from 0 to 0.6 are written in Table 5. It is clear that by increasing $\rho$, the transmitted power is increased due to higher favorable price deviation from the forecasted values, which results in higher power generation and gas consumption of zone one.

**Table 5** Output results for the opportunity schedule

| $\rho$ | $P_{iz}(MW)$ | $P_g(MW)$ | $S_k(10^6 Sm^3)$ |
| --- | --- | --- | --- |
| 0.0 | 33453 | 78451 | 40.198 |
| 0.1 | 33538 | 78536 | 40.208 |
| 0.2 | 33642 | 78639 | 40.237 |
| 0.3 | 33717 | 78715 | 40.256 |
| 0.4 | 33776 | 78778 | 40.265 |
| 0.5 | 33788 | 78790 | 40.272 |
| 0.6 | 33800 | 78802 | 40.279 |

The hourly opportunity schedule of the transmitted power, two selected generation units 3 and 11 as well as the natural gas suppliers of Peronnes and Voeren under different cost deviation factors $\rho = 0.1$ and $\rho = 0.6$ are presented in Figs. 7-9 to investigate the effect of the different cost targets on the scheduling, respectively. The results are gained for $B_w = (1 - 0.1)B_0 = \$381135$ and $B_w = (1 - 0.6)B_0 = \$169393$. Considering Fig. 7, in the case that the deviation factor $\rho$ increases from 0.1 to 0.6, the active power of another branch will increase if the transmitted power via one of the boundary branches decreases. At some hours such as *t*=6, the transmitted power from line 12 to 25 is constant at two cases $\rho = 0.1$ and $\rho = 0.6$. Meanwhile, the boundary branch 17 to 26 transmits more electrical power to the adjacent areas in case of $\rho = 0.6$. In other words, by increasing the cost deviation factor $\rho$, the total value of the active power transmitted from zone one to areas two and three is increased over the 24-hour study horizon. Hence, the revenue obtained from selling electricity is increased due to higher transmitted power. As seen in Fig. 8 for the higher $\rho$ the generation power increases because it is profitable to produce more power due to the higher positive price spikes.

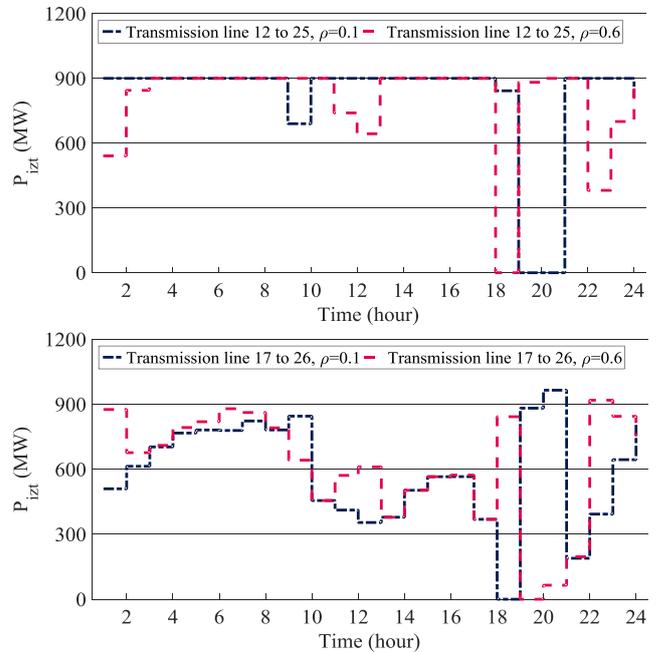

***Fig. 7.*** *Opportunity schedule of transmitted power for two different cost targets*

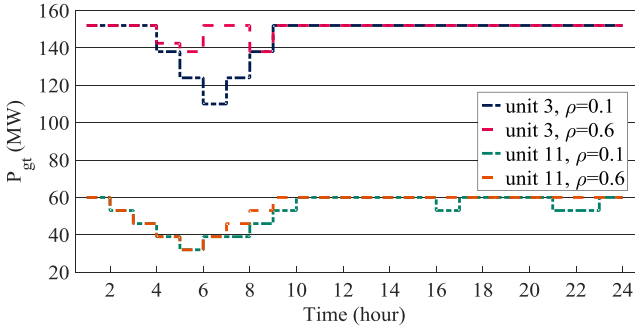

*Fig. 8. Opportunity schedule of units 3 and 11 for two different cost target.*

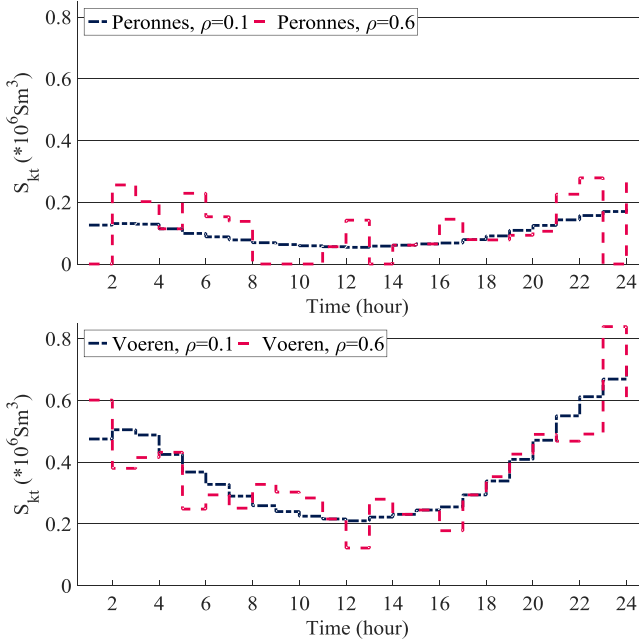

*Fig. 9. Opportunity schedule of gas suppliers for two different cost targets.*

According to Figs. 7-9, when the cost deviation factor $\rho$ increases, the optimum generation schedule of the power generation units and gas suppliers changes in a way that the electrical power transmitted through the boundaries 12 to 25 and 17 to 26 as well as the revenue obtained from selling energy increases. Therefore, the cost value, which is equal to the operation cost of the integrated system minus the revenue achieved from energy sale, decreases.

## 5. Conclusion

This paper studies the coordination of NG and electricity systems where gas-fired units link the electricity system and NG infrastructure. The objective is to minimize the cost of one specific zone while satisfying the constraints of the two interdependent systems, which can obtain revenue from selling power to its connected zones in short-term scheduling. The uncertainty of market price affects the economic operation of both NG and electricity systems of zone one; thus, the risk-based IGDT approach is employed to model the uncertainty of electricity price. If the future market price decreases within the maximum robustness horizon, the proposed robustness function states that the maximum cost of zone one is not higher than a specified cost. The defined opportunity function shows that the cost of zone one is lower than a given cost target by minimum increment within the opportunity region in the price market. For the risk-averse and opportunity-seeker decision makers, the robustness and opportunity results are respectively useful. The proposed method is tested on 20-node NG network and IEEE RTS 24-bus in zone one. The simulation results verify the capability of the proposed method.

**Acknowledgement**

The work done by Alireza Soroudi is supported by a research grant from Science Foundation Ireland (SFI) under the SFI Strategic Partnership Programme Grant No. SFI/15/SPP/E3125. The opinions, findings and conclusions or recommendations expressed in this material are those of the author(s) and do not necessarily reflect the views of the Science Foundation Ireland.

## 6. References

1. Chuan, H., Tianqi, L., Lei, W., and Shahidehpour, M., 'Robust Coordination of Interdependent Electricity and Natural Gas Systems in Day-Ahead Scheduling for Facilitating Volatile Renewable Generations Via Power-to-Gas Technology', *Journal of Modern Power Systems and Clean Energy*, 2017, 5, (3), pp. 375-388.
2. Li, T., Eremia, M., and Shahidehpour, M., 'Interdependency of Natural Gas Network and Power System Security', *IEEE Transactions on Power Systems*, 2008, 23, (4), pp. 1817-1824.
3. Qadrdan, M., Chaudry, M., Jenkins, N., Baruah, P., and Eyre, N., 'Impact of Transition to a Low Carbon Power System on the Gb Gas Network', *Applied Energy*, 2015, 151, pp. 1-12.
4. Alabdulwahab, A., Abusorrah, A., Zhang, X., and Shahidehpour, M., 'Coordination of Interdependent Natural Gas and Electricity Infrastructures for Firming the Variability of Wind Energy in Stochastic Day-Ahead Scheduling', *IEEE Transactions on Sustainable Energy*, 2015, 6, (2), pp. 606-615.
5. Correa-Posada, C.M. and Sanchez-Martin, P., 'Security-Constrained Optimal Power and Natural-Gas Flow', *IEEE Transactions on Power Systems*, 2014, 29, (4), pp. 1780-1787.
6. Quelhas, A., Gil, E., McCalley, J.D., and Ryan, S.M., 'A Multiperiod Generalized Network Flow Model of the Us Integrated Energy System: Part I—Model Description', *IEEE Transactions on Power Systems*, 2007, 22, (2), pp. 829-836.
7. Pantoš, M., 'Market-Based Congestion Management in Electric Power Systems with Increased Share of Natural Gas Dependent Power Plants', *Energy*, 2011, 36, (7), pp. 4244-4255.
8. Zhang, X., Shahidehpour, M., Alabdulwahab, A., and Abusorrah, A., 'Hourly Electricity Demand Response in the Stochastic Day-Ahead Scheduling of Coordinated Electricity and Natural Gas Networks', *IEEE Transactions on Power Systems*, 2016, 31, (1), pp. 592-601.
9. Unsihuay, C., Marangon-Lima, J., and de Souza, A.Z., 'Short-Term Operation Planning of Integrated Hydrothermal and Natural Gas Systems', in, *Power Tech, 2007 IEEE Lausanne*, (IEEE, 2007)

10. Liu, C., Shahidehpour, M., and Wang, J., 'Application of Augmented Lagrangian Relaxation to Coordinated Scheduling of Interdependent Hydrothermal Power and Natural Gas Systems', *IET generation, transmission & distribution*, 2010, 4, (12), pp. 1314-1325.
11. Martinez-Mares, A. and Fuerte-Esquivel, C.R., 'A Unified Gas and Power Flow Analysis in Natural Gas and Electricity Coupled Networks', *IEEE Transactions on Power Systems*, 2012, 27, (4), pp. 2156-2166.
12. Liu, C., Shahidehpour, M., Fu, Y., and Li, Z., 'Security-Constrained Unit Commitment with Natural Gas Transmission Constraints', *IEEE Transactions on Power Systems*, 2009, 24, (3), pp. 1523-1536.
13. Munoz, J., Jimenez-Redondo, N., Perez-Ruiz, J., and Barquin, J., 'Natural Gas Network Modeling for Power Systems Reliability Studies', in, *Power Tech Conference Proceedings, 2003 IEEE Bologna*, (IEEE, 2003)
14. Qadrdan, M., Chaudry, M., Wu, J., Jenkins, N., and Ekanayake, J., 'Impact of a Large Penetration of Wind Generation on the Gb Gas Network', *Energy Policy*, 2010, 38, (10), pp. 5684-5695.
15. Kamalinia, S., Wu, L., and Shahidehpour, M., 'Stochastic Midterm Coordination of Hydro and Natural Gas Flexibilities for Wind Energy Integration', *IEEE Transactions on Sustainable Energy*, 2014, 5, (4), pp. 1070-1079.
16. Qadrdan, M., Wu, J., Jenkins, N., and Ekanayake, J., 'Operating Strategies for a Gb Integrated Gas and Electricity Network Considering the Uncertainty in Wind Power Forecasts', *IEEE Transactions on Sustainable Energy*, 2014, 5, (1), pp. 128-138.
17. Bai, L., Li, F., Cui, H., Jiang, T., Sun, H., and Zhu, J., 'Interval Optimization Based Operating Strategy for Gas-Electricity Integrated Energy Systems Considering Demand Response and Wind Uncertainty', *Applied Energy*, 2016, 167, pp. 270-279.
18. Soroudi, A. and Amraee, T., 'Decision Making under Uncertainty in Energy Systems: State of the Art', *Renewable and Sustainable Energy Reviews*, 2013, 28, pp. 376-384.
19. Soroudi, A., Rabiee, A., and Keane, A., 'Information Gap Decision Theory Approach to Deal with Wind Power Uncertainty in Unit Commitment', *Electric Power Systems Research*, 2017, 145, pp. 137-148.
20. Ben-Haim, Y., *Info-Gap Decision Theory: Decisions under Severe Uncertainty*, (Academic Press, 2006)
21. Mohammadi-Ivatloo, B., Zareipour, H., Amjady, N., and Ehsan, M., 'Application of Information-Gap Decision Theory to Risk-Constrained Self-Scheduling of Gencos', *IEEE Transactions on Power Systems*, 2013, 28, (2), pp. 1093-1102.
22. Soroudi, A. and Ehsan, M., 'IGDT Based Robust Decision Making Tool for Dnos in Load Procurement under Severe Uncertainty', *IEEE Transactions on Smart Grid*, 2013, 4, (2), pp. 886-895.
23. Shafiee, S., Zareipour, H., Knight, A.M., Amjady, N., and Mohammadi-Ivatloo, B., 'Risk-Constrained Bidding and Offering Strategy for a Merchant Compressed Air Energy Storage Plant', *IEEE Transactions on Power Systems*, 2017, 32, (2), pp. 946-957.
24. He, Y., Shahidehpour, M., Li, Z., Guo, C., and Zhu, B., 'Robust Constrained Operation of Integrated Electricity-Natural Gas System Considering Distributed Natural Gas Storage', *IEEE Transactions on Sustainable Energy*, 2018, 9, (3), pp. 1061-1071.
25. Nojavan, S., Ghesmati, H., and Zare, K., 'Robust Optimal Offering Strategy of Large Consumer Using Igdt Considering Demand Response Programs', *Electric Power Systems Research*, 2016, 130, pp. 46-58.
26. Zareipour, H., Janjani, A., Leung, H., Motamedi, A., and Schellenberg, A., 'Classification of Future Electricity Market Prices', *IEEE Transactions on Power Systems*, 2011, 26, (1), pp. 165-173.
27. He, Y., Yan, M., Shahidehpour, M., Li, Z., Guo, C., Wu, L., and Ding, Y., 'Decentralized Optimization of Multi-Area Electricity-Natural Gas Flows Based on Cone Reformulation', *IEEE Transactions on Power Systems*, 2018, 33, (4), pp. 4531-4542.
28. Rosenthal, E., 'Gams-a User's Guide', in, *GAMS Development Corporation*, (Citeseer, 2008)
29. Brooke, A.K. and D Meeraus, A., Gams Release 2.25; a User's Guide', (GAMS Development Corporation, Washington, DC (EUA), 1996)
30. Soroudi, A., Power System Optimization Modeling in Gams', (Springer, 2017)